\documentclass[preprint, aps]{revtex4}
\usepackage{graphicx}
\usepackage{dcolumn}
\usepackage{bm}
\usepackage{slashbox}
\usepackage{epsfig}
\usepackage{slashbox}
\usepackage{subfigure}

\begin{document}

\title{Deconfinement phase transition in holographic QCD with matter}

\author{Youngman Kim}
 \affiliation{School of Physics, Korea Institute for Advanced Study , Seoul 130-722, Korea}
 \email{chunboo81@kias.re.kr}

\author{Bum-Hoon Lee}
 \email{bhl@sogang.ac.kr}
 \affiliation{Department of Physics, Sogang University, Seoul, Korea 121-742}
 \affiliation{CQUeST, Sogang University, Seoul, Korea 121-742}

\author{Siyoung Nam}
 \email{stringphy@gmail.com}
 \affiliation{CQUeST, Sogang University, Seoul, Korea 121-742}

\author{Chanyong Park}
 \email{cyong21@sogang.ac.kr}
 \affiliation{CQUeST, Sogang University, Seoul, Korea 121-742}

\author{Sang-Jin Sin}
 \email{sjsin@hanyang.ac.kr}
 \affiliation{Department of physics, BK21 Program Division, Hanyang University, Seoul 133-791, Korea}

\begin{abstract}
In the framework of a holographic QCD approach we study an
influence of matters on the deconfinement temperature, $T_c$. We
first consider quark flavor number ($N_f$) dependence of $T_c$. We
observe that $T_c$ decreases with $N_f$, which is consistent with
a lattice QCD result. We also delve into how the quark number
density $\rho_q$ affects the value of $T_c$. We find that
$T_c$ drops with increasing $\rho_q$. In both cases, we confirm that the
contributions from quarks are suppressed by $1/N_c$, as it should be, compared to the ones from a
 gravitational action (pure Yang-Mills).
\end{abstract}

\maketitle

\section{Introduction}
To understand QCD phase structure at finite temperature and/or density has been an fascinating theme in hadron physics.
 Some of studies so far are basically  based on phenomenological models or effective field theories of QCD such
  as chiral perturbation theory (ChPT) and Nambu-Jona-Lasinio (NJL) model.  Lattice QCD has been a powerful tool
   for QCD phase diagram at finite temperature and, recently, at finite density.

Recently interesting developments in AdS/CFT~\cite{adscft} to study strongly interacting
system such as QCD, which goes under the name of AdS/QCD, have been made. Confinement is realized
 with an infrared (IR) cut off $z_{IR}$ in AdS space~\cite{polchinski},
  and flavors are introduced by adding extra probe branes~\cite{karch}.
   More phenomenological approaches were also suggested to construct a holographic model
   dual to QCD, for example, ~\cite{EKSS,PR,Brodsky}. The deconfinement temperature is estimated in
    a Hawking-Page type transition analysis in cutoff AdS space in ~\cite{Herzog:2006ra}.
    In this analysis, the contribution from mesons (quarks) are not considered,
    since they are suppressed by $1/N_c$ compared to the gravitational
    part, and consequently, $T_c$ may correspond to that of pure
    Yang-Mills. The quark (baryon) chemical potential is introduced
    through  AdS/CFT to study physics of dense matter~\cite{dAdSQCD}.

In the present work, we consider contributions from mesons, motivated
largely by lattice QCD results, to see flavor $N_f$ and quark number
density
dependence of $T_c$. As it is well know from many lattice QCD results
so far,
 the details of the transition strongly depends on the number of quark
 flavors.
 The transition temperature with no dynamical quarks is roughly
 $270~{\rm MeV}$,
 for instance see~\cite{Karsch}, while with dynamical quarks it is
 around $170~{\rm MeV}$~\cite{KLP,Aoki}.
 In addition, $T_c$ will decreases with increasing $N_f$~\cite{KLP}.
 Apart from the temperature axis in the QCD phase diagram, lattice QCD
 is now investigating
 the QCD equation of state  at nonzero chemical potential~\cite{FK,AEH}. In \cite{FK}, it is shown that the transition temperature decreases with the baryon chemical potential up to ~$1~{\rm GeV}$.
Finally, we remark that $T_c$ in the present work denotes the critical temperature of the deconfinement (first order) phase transition, while at low density there is a cross over as observed in lattice QCD ~\cite{FK,Aoki,Bernard}. We attribute this difference to the large $N_c$ nature of the Hawking-Page type transition.

\section{A Hawking-Page type transition with quark flavors}
A study~\cite{Herzog:2006ra} based on a Hawking-Page type transition in the AdS/QCD models calculated the deconfinement temperature. In the analysis, the contribution from mesons are not considered, since they are suppressed by $1/N_c$ compared to the gravitational part: the gravitational coupling scales as $\kappa \approx g_s \approx 1/N_c$, and the contribution from the mesons scales only as $N_c$. Here we briefly summarize the analysis of ~\cite{Herzog:2006ra} done in the hard wall model~\cite{EKSS, PR}. In the hard wall model, the AdS space is compactified such that $z_0<z<z_{IR}$, where
$z_0\rightarrow 0$. The value of $z_{IR}$ is fixed by the rho-meson mass ($m_\rho$) at zero temperature: $m_\rho (\simeq 770~{\rm MeV})\simeq3\pi/(4z_{IR})$$\rightarrow$ $1/z_{IR}\simeq 320~{\rm
MeV}$\cite{EKSS, PR}.

The Euclidean gravitational action given by
\begin{equation} \label{action1}
S_{grav} ~=~ -\frac{1}{2\kappa^2} \int d^5x \sqrt{g}\left(\textrm{R}+\frac{12}{L^2}\right)
\end{equation}
where $\kappa^2 = 8\pi G_5$ and $L$ is the length scale of the $AdS_5$, there are two relevant solutions for the equations of motion derived from the above action.The one is cut-off thermal AdS(tAdS) with the line element
\begin{equation}
ds^2=\frac{L^2}{z^2}\left(d\tau^2+dz^2+d\vec{x}^2_3\right),
\end{equation}
where the radial coordinate runs from the boundary of tAdS space $z=0$ to the
cut-off $z_{IR}$, which corresponds to an infrared cut-off in energies proportional to $1/z_{IR}$ from the point of view of the boundary dual theory. The other solution is cut-off AdS black hole(AdSBH) with the line element
\begin{equation}
ds^2=\frac{L^2}{z^2}\left(f(z)d\tau^2+\frac{dz^2}{f(z)}+d\vec{x}^2_3\right)
\end{equation}
where $f(z)=1-(z/z_h)^4$ and $z_h$ is the horizon of the black hole. Note that there will be no cut-off in this space for $z_{IR}<z_h$. The Hawking temperature of the black hole solution is $T = 1/(\pi z_h)$ which is given by regularizing the metric near the horizon. In the tAdS case, the periodicity in the Euclidean time-direction is fixed by comparing two geometries at an UV cut-off $\epsilon$ where the periodicity of the time-direction in both cases is locally the same. Then, the time periodicity of tAdS is given by
\begin{equation}
\beta = \pi z_h \sqrt{f(\epsilon)}.
\end{equation}
Now we calculate the action density $V$, which is defined by the action divided by the common volume factor of $R^3$. The regularized action density of the tAdS is given by
\begin{equation}
V_1(\epsilon) = \frac{4L^3}{\kappa^2} \int^{\beta '}_{0} d\tau \int^{z_{IR}}_{\epsilon}\frac{dz}{z^5}\, ,
\end{equation}
and that of the AdSBH is given by
\begin{equation}
V_2(\epsilon) = \frac{4L^3}{\kappa^2}\int^{\pi z_h}_{0} d\tau \int^{\bar{z}}_{\epsilon}\frac{dz}{z^5}
\end{equation}
where $\bar{z} = min(z_{IR},z_h)$.
Then, the difference of the regularized actions is given by
\begin{equation}
\Delta V_g = \lim_{\epsilon\rightarrow 0}\left[V_2(\epsilon) -V_1(\epsilon)\right]= \left\{\begin{array}{ll} \frac{L^3 \pi z_h}{\kappa^2}
\frac{1}{2z_h^4} & z_{IR} < z_h\\ \\
\frac{L^3 \pi z_h}{\kappa^2} \left( \frac{1}{z_{IR}^4} - \frac{1}{2z_h^4}\right)
& z_{IR} > z_h.\end{array}
\right.
\end{equation}
This is the result of \cite{Herzog:2006ra} in the hard wall model. When $\Delta V_g$ is positive(negative), tAdS  (the black hole) is stable. Thus, at $\Delta V_g=0$ there exists a Hawking-Page transition. In the first case $z_{IR} < z_h$, there is no Hawking-Page transition and the thermal AdS is always stable. In the second case $z_{IR} > z_h$, the Hawking-Page transition occurs at
\begin{equation}    \label{tempads}
T_0 = 2^{1/4}/(\pi z_{IR})
\end{equation}
and at low temperature $T < T_0$ (at high temperature $T > T_0$) the
thermal AdS (the AdS black hole) geometry becomes a dominant background.

Now, we include the bulk matter into the theory. The action for mesons is given by
\begin{equation} \label{action2}
S_{matter} ~=~  M_5 \int d^5x \sqrt{g}~ \textrm{Tr} \left[\frac{1}{2}|D_\mu \Phi|^2
+ \frac{1}{2} M_{\Phi}^2|\Phi^2| +\frac{1}{4}\left(F_{L}^2 +F_{R}^2\right)\right]
\end{equation}
where $L^2 M_{\Phi}^2=-3$,
$D_\mu \Phi=\partial_\mu \Phi +i A_{L\mu} \Phi -i\Phi A_{R\mu}$, $A_{L,R} = A_{L,R}^a t^a$ and $F_{\mu\nu}=\partial_\mu A_\nu - \partial_\nu A_\mu +i[A_\mu,A_\nu]$.
The vector fields and the axial vector fields are defined by $V=(A_L +A_R)/\sqrt{2}$ and $A=(A_L -A_R)/\sqrt{2}$ respectively. For later convenience, we present some relations here:
\begin{equation}
\frac{1}{\kappa^2}=\frac{1}{8\pi G_5}, ~~~~ \frac{1}{G_5}=\frac{32N^2_c}{\pi L^3},
~~ ~~ \textrm{and}~~~~M_5=\frac{N_c}{12\pi^2 L},
\end{equation}
where the second relation comes from Ref. \cite{Csaki:2006ji}. In this section, we study the quark flavor number dependence of the critical temperature of the Hawking-Page type transition at zero density. Here we assume that there is no condensate in the vector and axial-vector channel, and so we turn off the vector gauge fields temporarily until next section. On thermal AdS background, the solution of the equation of motion for a scalar field is given by
\begin{equation}
v(z) = a z + b z^3 \, ,
\end{equation}
where $v(z)=<\Phi>$.
According to the AdS/CFT correspondence, the coefficient $a$ is a mass of the boundary quark and $b$ corresponds to chiral condensate $<\bar qq>$. When considering the AdS black hole background, the boundary theory becomes a finite temperature field theory. The equation of motion of the scalar field is
\begin{equation}
\left[\partial_z^2 -\frac{4-f}{zf}\partial_z +\frac{3}{z^2f}\right]v(z) =0
\end{equation}
and the solution of the equation is given by~\cite{GY, Kim:2006ut}
\begin{equation}    \label{sinbh}
v(z) = a \cdot _2F_1\left(\frac{1}{4}, \frac{1}{4}, \frac{1}{2},
\frac{z^4}{z_h^4}\right)  z + \tilde b \cdot _2F_1\left(\frac{3}{4}, \frac{3}{4},
\frac{3}{2}, \frac{z^4}{z_h^4}  \right) z^3 .
\end{equation}
For simplicity, we take the chiral limit where the quark mass is zero, $a=0$.
We assume that chiral symmetry restoration and the deconfinement take place at the same temperature, and so the chiral condensate is zero in the deconfined phase. Since AdS black hole background corresponds to a deconfined phase of the boundary theory,  we have $\tilde b=0$. At low temperature described by the thermal AdS background, the boundary theory corresponds to a confined phase, so $b\neq  0$. Therefore, the scalar field solutions in two backgrounds are given by $v_{tAdS}\simeq b z^3$ and $v_{BH}\simeq 0$.

Now we calculate the matter contribution to the Hawking-Page transition. For the thermal AdS, the scalar field contribution to the action is
\begin{equation}
V_{1m} = \frac{L^3\pi z_h}{2}\cdot 3M_5 N_f b^2z^2_{IR}.
\end{equation}
In the black hole background, since $v_{BH}\simeq 0$ there is no contribution of the scalar field, i.e. $V_{2m}=0$. Therefore, the values of $\Delta V_m$
of the matter part for $z_{IR} < z_h$ and $z_{IR} > z_h$ are the same :
\begin{equation}
\Delta V_m = -\frac{L^3 \pi z_h}{\kappa^2}\cdot \frac{b_t^2 L^2 N_f z_{IR}^2}{32N_c}.
\end{equation}
Note that the parameter $b$ is given by
\begin{equation}
a \simeq 0 ~~\textrm{and}~~ b \simeq \frac{\xi}{L z_{IR}^3}
\end{equation}
where $\xi \simeq4$  \cite{PR}.
Finally, the total difference of the action in each background becomes
\begin{equation}
\Delta V = \left\{\begin{array}{ll}\frac{L^3 \pi z_h}{2\kappa^2}\left[ \frac{1}{z_h^4}
- \left(\frac{N_f}{N_c}\right)\frac{\xi^2}{16 z_{IR}^4}\right] & z_{IR} < z_h\\ \\
\frac{L^3 \pi z_h}{2\kappa^2} \left[ \frac{2}{z_{IR}^4} - \frac{1}{z_h^4}
- \left(\frac{N_f}{N_c}\right)\frac{\xi^2}{16 z_{IR}^4}\right] & z_{IR} > z_h.
\end{array} \right.
\end{equation}
If we define the critical temperature of the pure $AdS_5$
 gravity (pure Yang-Mills) as $T_0  =\frac{1}{\pi z_h}
 =\frac{2^{1/4}}{\pi z_{IR}}$
 following (\ref{tempads}) the critical temperatures modified by mesons is given by
\begin{equation}
T = \left\{\begin{array}{ll} {T}_0 \left(\frac{\xi^2}{32}
\frac{N_f}{N_c}\right)^{\frac{1}{4}} & z_{IR} <z_h\\ \\
{T}_0 \left(1
-\frac{\xi^2}{32}\frac{N_f}{N_c}\right)^{\frac{1}{4}} & z_{IR}>z_h.\end{array}\right.
\end{equation}
In the first case $z_{IR} <z_h$, a Hawking-Page type transition
 occurs, which  seems unphysical, since it implies that there is a
 phase transition at low temperature. Unfortunately, we have no clear
 understanding on this phase transition. If we include, however, the
 back-reaction of the matter field on the background and work on a
 deformed AdS metric, then the unphysical phase transition may
 disappear.

In the second case $z_{IR}>z_h$, at high (low) temperature the AdS black hole (the thermal AdS) is again stable, but due to the matter field, the critical temperature is smaller than that of the pure AdS gravity theory, which  is consistent with an observation made in~\cite{KLP}.

\section{A Hawking-Page type transitions at finite density}
In this section, we consider the quark number density dependence on the deconfinement temperature. In QCD, quark chemical potential introduced as $\mu_q\bar\psi\gamma_0\psi$, and so according to an AdS/CFT dictionary
we need to introduce a bulk U(1) field in AdS$_5$ whose boundary value is $\mu_q$. To this end, we follow~\cite{DH} and generalize the symmetry to U$(N_f)\times $U$(N_f)$. The equation of motion for the time component of the U(1) vector field is given by
\begin{equation}
\partial_z\left[\frac{1}{z}\partial_z V_{\tau}(z)\right] = 0\, ,
\end{equation}
and a solution of the equation of motion is given by
\begin{equation}
V_{\tau} = c_1 +  c_2 z^2\, .
\end{equation}
Since the factors $g^{\tau\tau}g^{zz}$ in the equation of motion for tAdS and AdSBH backgrounds are same, we conclude that the equations of motion of both backgrounds are same and the corresponding forms of the solutions are also same. That is, we can use the same form of the solution $V_{\tau}$ in both background tAdS and AdSBH. According to the AdS/CFT correspondence, the coefficient of the non-normalizable term, $c_1$, is proportional to coupling with the dual operator of the boundary theory. Since the time component of the U(1) vector field is dual to the quark number current, $c_1$ must correspond to the quark chemical potential. Meanwhile, the coefficient of the normalizable term, $c_2$, corresponds to the expectation value of the dual operator so that $c_2$ is interpreted as the quark number density, $c_2=12\pi^2\rho_q/N_c$~\cite{DH}.

Following the same procedure used in the previous section, we arrive at
\begin{equation}
V_{v1} = \pi z_h M_5 N_f L^5 c_2^2 z_{IR}^2
\end{equation}
for the tAdS and
\begin{equation}
V_{v2}=\left \{\begin{array}{ll} \pi z_h M_5 N_f L^5 c_2^2 z_{h}^2 & z_h<z_{IR}\\
\\\pi z_h M_5 N_f L^5 c_2^2 z_{IR}^2 & z_h>z_{IR} \end{array} \right.
\end{equation}
for AdSBH. From these results, the differences of the action reads
\begin{equation}
\Delta V_{v}=\left\{\begin{array}{ll} -\pi z_h M_5 N_f L^5 c_2^2(z^2_{IR}- z_{h}^2)
& z_h<z_{IR}\\ \\0 & z_h>z_{IR} \end{array} \right.
\end{equation}
The final result for $z_h<z_{IR}$ is
\begin{equation}    \label{hp2}
\Delta V = \frac{L^3 \pi z_h}{\kappa^2}\left[ \frac{1}{z_{IR}^4} -\frac{1}{2z_h^4}
-\frac{L^4 N_f c^2_2}{48N_c}\left(z^2_{IR}- z_{h}^2\right)\right]
\end{equation}
When $\Delta V < 0$, the dominant geometry is a AdS black hole and the boundary theory corresponding to this geometry is in a confined phase. As in Ref. \cite{Herzog:2006ra}, in the pure gravity theory, the critical temperature
is given by
$T_0 = \frac{2^{1/4}}{\pi z_{IR}}$.
So when considering the quark density, using the relation $T = 1/(\pi z_h)$,
the critical temperature $T_c$ is determined by solving the following equation
\begin{equation} \label{eTc}
0~=~ T^4 - \frac{2}{\pi^4} \left( \frac{1}{z_{IR}^4}
-\frac{L^4 N_f }{48N_c} c^2_2 \left( z^2_{IR} - \frac{1}{\pi^2 T^2} \right) \right)\Bigg{|}_{T =T_c} .
\end{equation}
In (\ref{eTc}), the last term $z^2_{IR}- z_{h}^2$,
which is from quark number density, is positive, and so the critical temperature at finite density is always lower than that of the pure gravity theory. Now, we consider a case where $z_h\ll z_{IR}$ to see the quark number dependence of $T_c$ clearly. We note here that typically $z_h^2/z_{IR}^2\approx 0.6$.

\begin{figure}[!ht]
\begin{center}
\subfigure[] {\includegraphics[angle=0,
width=0.4\textwidth]{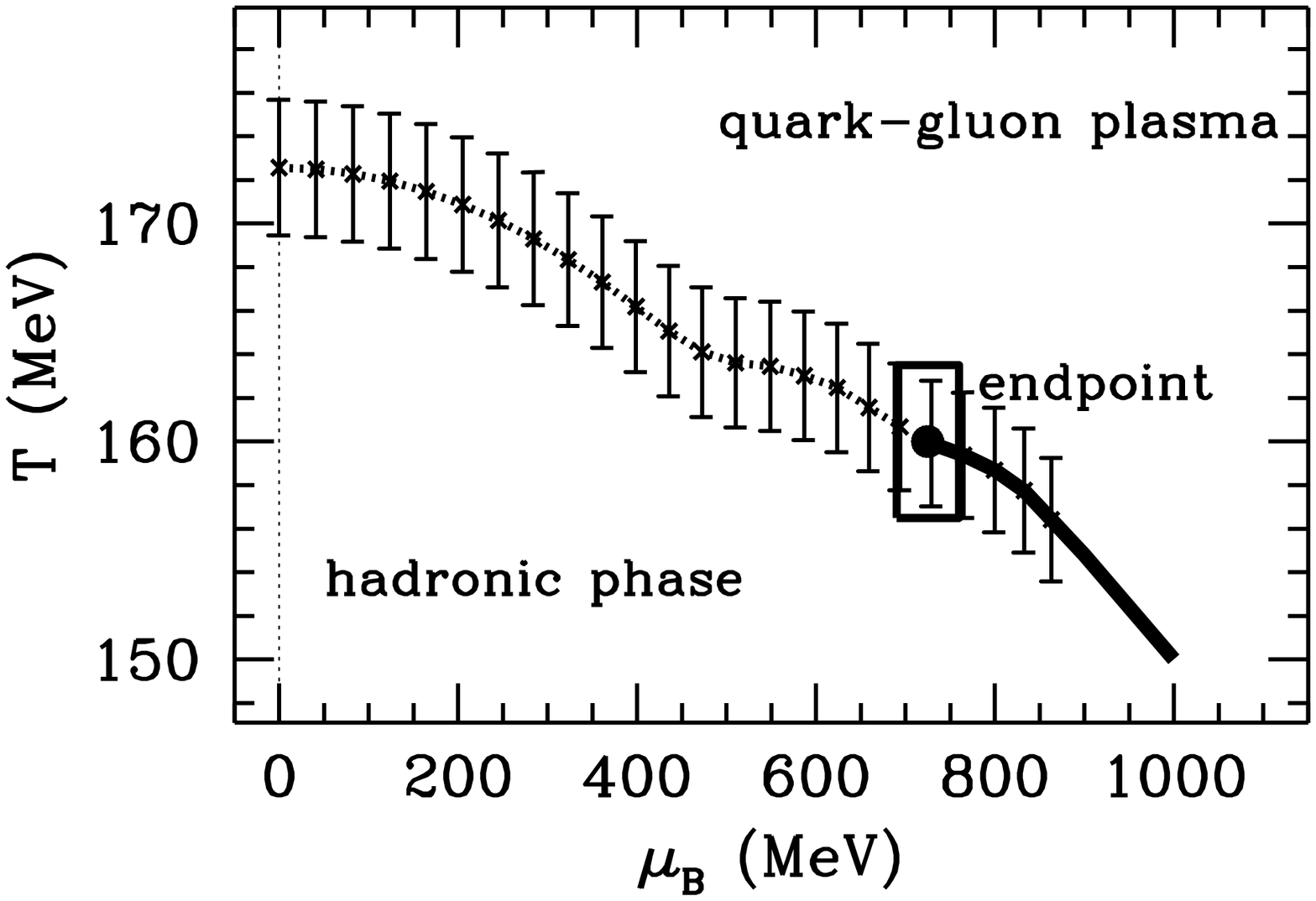} \label{fig:FK}}
\subfigure[] {\includegraphics[angle=0,
width=0.4\textwidth]{{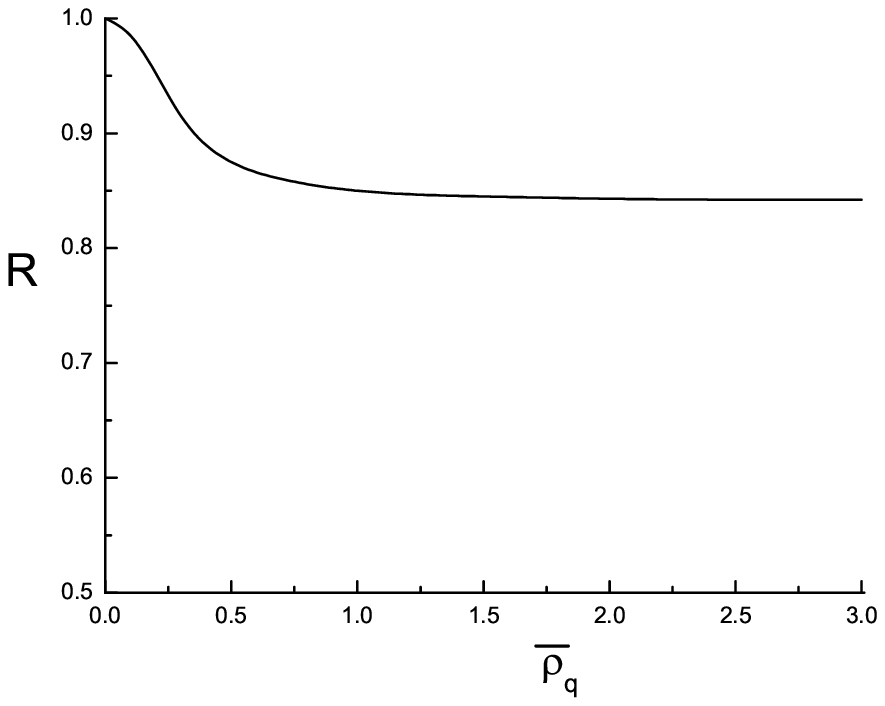}} \label{fig:HPt}}
 \caption{ (a) The transition temperature from lattice simulations~\cite{FK, FKtalk}. Here dotted line denotes the crossover and the solid line is for the first order phase transition. The figure is taken from \cite{FKtalk}. (b) Density dependence of $T_c$ from Eq.~(\ref{eTc}). Here $R\equiv T_c(\rho_q)/T_0$ and $\bar\rho_q =\rho_q z_{IR}^3$. \label{TcF} }
\end{center}
\end{figure}

In this case, the critical temperature is given by
\begin{equation} \label{Tc}
T_c (\rho_q) = \frac{2^{1/4}}{\pi} \left( \frac{1}{z_{IR}^4}
-\frac{L^4 N_f z^2_{IR}}{48N_c}  c^2_2 \right)^{1/4},
\end{equation}
which is a consistent result with the lattice QCD data.
Note here again that $c_2\sim \rho_q$~\cite{DH}. In Fig.~\ref{TcF}, we plot Eq.~(\ref{eTc}), together with the lattice result~\cite{FK, FKtalk},
where $R\equiv T_c(\rho_q)/T_0$ and $\bar\rho_q =\rho_q z_{IR}^3$.
Our result in Fig.~\ref{TcF} is consistent with lattice QCD results at low
density,~{\it i.e}, see~\cite{FK}.

Finally, we combine the results obtained in the previous and in this sections. The final form of $\Delta V$ is given by
\begin{equation}
\Delta V = \left\{
\begin{array}{ll}
\frac{L^3 \pi z_h}{2\kappa^2}\left[ \frac{1}{z_h^4} - \left(\frac{N_f}{N_c}\right)
\frac{\xi^2}{16 z_{IR}^4}\right] & z_{IR} < z_h\\ \\ \frac{L^3 \pi z_h}{2\kappa^2}
\left[ \frac{2}{z_{IR}^4} - \frac{1}{z_h^4}- \left(\frac{N_f}{N_c}\right)
\frac{\xi^2}{16 z_{IR}^4} -\frac{L^4 N_f c^2_2}{48N_c}\left(z^2_{IR}- z_{h}^2\right)
\right] & z_{IR} > z_h. \end{array} \right.
\end{equation}

\section{Discussion}
We have studied the effects of matters on the deconfinement transition in the context of a Hawking-Page type analysis. We observed, as it should be, that the corrections from mesons to the deconfinement temperature are suppressed by $1/N_c$. We found that $T_c$ decreases with the number of quark flavor $N_f$. At finite density, we obtained the density dependence of $T_c$, and it shows a similar behavior calculated in lattice QCD.

We remark here that, as it is well known, in the presence of the dynamical quarks the Polyakov loop is no longer a good order parameter to describe the deconfinement transition~\cite{BU}. In lattice QCD, however, it has been established that the Polyakov loop could be served as an approximate order parameter for the deconfinement transition. While, one of the key quantities that connects the Hawking-Page transition to the deconfinement is the Polyakov loop. In the present work, we assume that the the gravity description of confinement/deconfinement through the Hawking-Page transition is still viable with dynamical quarks. It may be interesting to see how the expectation value of the Polyakov loop behaves in the gravity side, when the dynamical quarks in the fundamental representation are present.

\vskip 1cm
\begin{acknowledgments}
The work of Bum-Hoon Lee, Siyoung Nam and Chanyong Park was supported by the Science Research Center Program of the Korea Science and Engineering Foundation through the Center for Quantum Spacetime(CQUeST) of Sogang University with grant number R11 - 2005 - 021.
\end{acknowledgments}


\end{document}